\documentclass[10pt,twocolumn]{IEEEtran}

\usepackage{cite}
\usepackage{amssymb}
\usepackage{amsmath}
\usepackage{dblfloatfix}
\usepackage{amsthm}
\usepackage{graphicx}
\graphicspath{{./graphics/}{.}}
\usepackage{color}
\usepackage{multirow}

\newcommand{\hd}{\hdots}
\newcommand{\hs}[1]{\hspace{-#1 ex}}
\long\def\symbolfootnote[#1]#2{\begingroup
\def\thefootnote{\fnsymbol{footnote}}\footnote[#1]{#2}\endgroup}
\newcommand{\Exp}[1]{\mathbb{E}\left\{#1\right\}}

\newcommand{\bsm}[1]{{\boldsymbol #1}}

\IEEEoverridecommandlockouts
\begin{document}

\title {Exact Cramer-Rao Bounds for Semi-blind Channel Estimation in Amplify-and-Forward Two-Way Relay Networks employing Square QAM Modulation}

\author{Saeed Abdallah and Ioannis N. Psaromiligkos*\thanks{The authors are with McGill University, Department of Electrical and Computer Engineering, 3480 University Street, Montreal, Quebec, H3A 2A7, Canada. Email: saeed.abdallah@mail.mcgill.ca; yannis@ece.mcgill.ca, phone: +1 (514) 398-2465, fax: +1 (514) 398-4470.}\thanks{*Corresponding author.}}

\maketitle
\begin{abstract}
In this paper, we derive the exact Cramer-Rao bounds (CRBs) for semi-blind channel estimation in amplify-and-forward two-way relay networks employing square QAM modulation. The derived bounds are used to show that the semi-blind approach, which exploits both the transmitted pilots and transmitted data symbols, can provide substantial improvements in estimation accuracy over the training-based approach which only uses pilot symbols to estimate the channel parameters. We also derive the more tractable modified CRB which accurately approximates the exact CRB at high SNR for low modulation orders. 

\begin{center}\textbf{\textit{Index Terms}:} Amplify and Forward, Cramer-Rao bound, Semi-blind Channel Estimation, Square QAM Modulation, Two-way Relays.\end{center}
\end{abstract}

\section{Introduction}

Amplify-and-forward (AF) two-way relay networks (TWRNs)~\cite{rankov07} have recently received a lot of attention as a spectrally efficient approach for bidirectional communication between two terminals. In previous works on TWRNs, it is often assumed that the channel parameters are perfectly known at the terminals. In reality, however, these parameters have to be estimated, and an estimation error is always incurred. Since part of the channel information is required for self-interference cancellation at each terminal, the presence of estimation error means that self-interference cannot be completely cancelled, and the residual interference reduces the overall performance of the system. 

The problem of estimating the channel in AF TWRNs has been addressed in a number of recent works~\cite{gao2009optimal,feifei_power,tensor_based,gao2009channel,joint_CFO,joint_detection,abdallah2012,partially_blind}. Most of these works adopt the pilot-based approach to channel estimation (c.f.~\cite{gao2009optimal}), which imposes an additional burden on the system and reduces the overall spectral efficiency. It is therefore important to investigate the application of semi-blind channel estimation which mitigates this burden by exploiting both pilots and the transmitted data to estimate the channel, or blind channel estimation which completely avoids this burden by relying only on the transmitted data for channel estimation. Researchers have recently started investigating such approaches~\cite{joint_detection,abdallah2012} in order to obtain better tradeoffs between estimation accuracy and spectral efficiency than the pilot-based approach. In~\cite{joint_detection}, a semi-blind approach which jointly estimates the channel and detects the transmitted data for MIMO-OFDM TWRNs was developed using the expectation conditional maximization (ECM) algorithm. In~\cite{abdallah2012}, a blind channel estimation algorithm was proposed for TWRNs employing constant modulus (CM) signalling in the form of $M$PSK modulation. 

In order to evaluate the potential of semi-blind and blind approaches in the AF TWRN context and to provide a benchmark on the performance of such algorithms, it is very useful to know the Cramer-Rao bound (CRB) on achievable estimation accuracy. This bound is more easily obtained for the pilot-based approach, as has been done in~\cite{gao2009optimal}. However, for semi-blind and blind approaches, the task of deriving the CRB is more challenging because of the complicated form of the likelihood function when the statistics of the data symbols are taken into account. In~\cite{abdallah2012} and~\cite{partially_blind}, the derivation of the blind CRB for the case of CM signalling was simplified by ignoring the statistics of the data symbols and instead treating them as deterministic unknowns. To the best of our knowledge, however, the exact CRBs that take into account the statistics of the data have not been derived before for AF TWRNs. In this work, we fill this gap and we derive the exact CRBs for semi-blind channel estimation in AF TWRNs employing square QAM modulation. Due to its bandwidth efficiency, this class of modulation schemes is very important for today's high datarate applications. The derived bounds also cover the blind and the pilot-based scenarios as special cases. We use the derived bounds to explore the potential of the semi-blind and blind approaches and show that substantial performance gains over the pilot-based approach can be achieved.

The rest of the paper is organized as follows. In Section~\ref{system_model} we present the system model. Our derivation of the CRB is presented in Section~\ref{CRB_section}. Simulation results are presented in~\ref{simulations}. Finally, our conclusions are in Section~\ref{conclusions}.

\begin{figure}[t]
\centering
\includegraphics[width=3.5in]{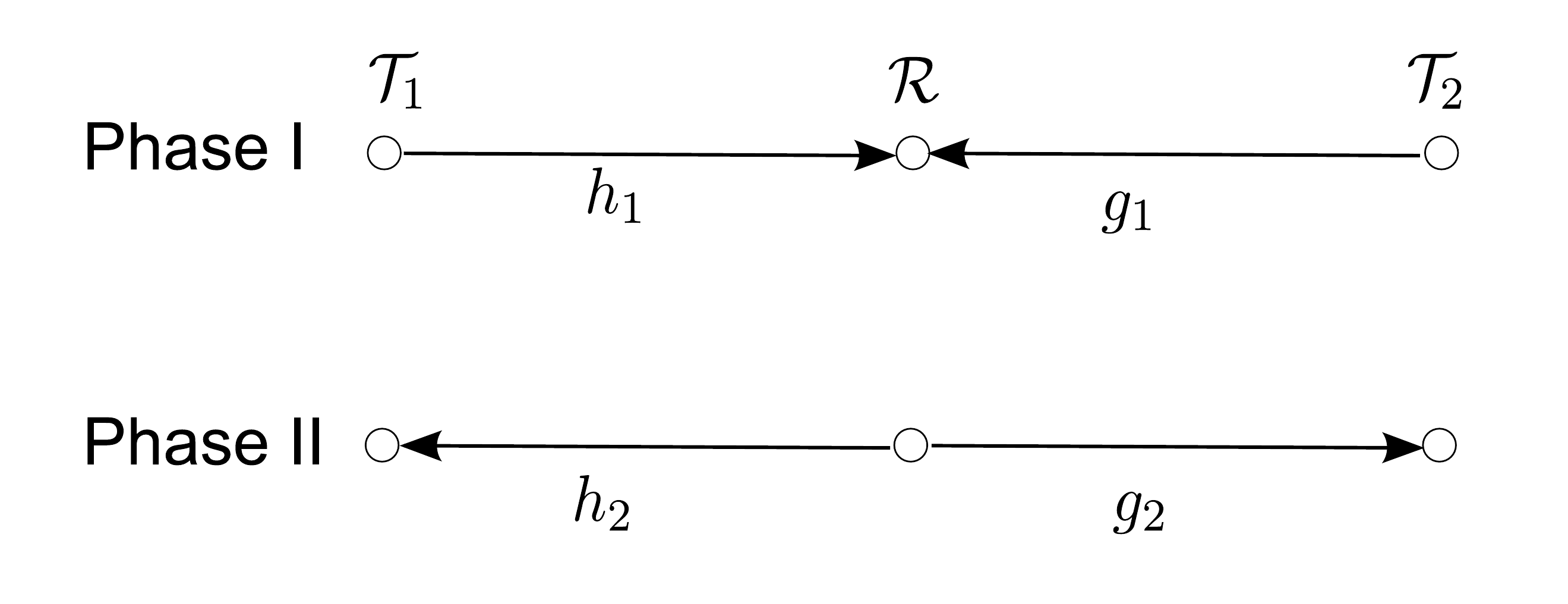}
\caption{The two-way relay network with two source terminals and one relay node.}
\label{two_way_nonreciprocal}
\end{figure}

\section{System Model}
\label{system_model}

We consider the half-duplex TWRN with two source nodes, $\mathcal{T}_1$ and $\mathcal{T}_2$, and a single relaying node $\mathcal{R}$, shown in Fig.~\ref{two_way_nonreciprocal}. The network operates in quasi-static flat-fading channel conditions. Each transmission period is divided into two phases. In the first phase, $\mathcal{T}_1$ and $\mathcal{T}_2$ simultaneously transmit to $\mathcal{R}$, and in the second phase $\mathcal{R}$ broadcasts an amplified version of the received signal to both terminals. The semi-blind approach employs both pilot symbols and data symbols to estimate the channel parameters. More specifically, prior to transmitting data symbols, each terminal transmits a block of $L$ pilot symbols. We denote by $\bsm{t}_1\triangleq[t_{11},\hd,t_{1L}]^{T}$ and $\bsm{t}_2\triangleq[t_{21},\hd,t_{2L}]$ the vectors containing the pilot symbols transmitted by $\mathcal{T}_1$ and $\mathcal{T}_2$, respectively. The received signal vector at $\mathcal{R}$ during the training period is $\bsm{r}_{t}=h_1\bsm{t}_1+g_1\bsm{t}_2+\bsm{\omega}$, where $h_1$ and $g_1$ are the complex coefficients of the flat-fading channels $\mathcal{T}_1\rightarrow\mathcal{R}$ and $\mathcal{T}_2\rightarrow\mathcal{R}$, respectively, and $\bsm{\omega}$ is the circular complex white Gaussian noise with mean zero and covariance $\sigma^2\bsm{I}$ $($denoted as $\mathcal{CCN}(0,\sigma^2\bsm{I}))$. The relay broadcasts $A\bsm{r}_t$, where $A>0$ is the amplification factor. The corresponding received signal vector at terminal $\mathcal{T}_1$ is 
\begin{equation}
\bsm{z}_t=Ah_1h_2\bsm{t}_1+Ag_1h_2\bsm{t}_2+Ah_2\bsm{\omega}+\bsm{\omega}_1
\end{equation}
where $\bsm{\omega}_1$ is also $\mathcal{CCN}(0,\sigma^2\bsm{I})$.

After transmitting the $L$ pilots, $\mathcal{T}_1$ and $\mathcal{T}_2$ transmit $N$ data symbols each. We denote by $\bsm{s}_1\triangleq[s_{11},\hd,s_{1N}]^{T}$ and $\bsm{s}_2\triangleq[s_{21},\hd,s_{2N}]^{T}$ the transmitted data symbol vectors of $\mathcal{T}_1$ and $\mathcal{T}_2$, respectively. The received signal vector at $\mathcal{R}$ is $\bsm{r}=h_1\bsm{s}_1+g_1\bsm{s}_2+\bsm{n}$, and the corresponding received signal vector at $\mathcal{T}_1$ is
\begin{equation}
\bsm{z}=Ah_1h_2\bsm{s}_1+Ag_1h_2\bsm{s}_2+Ah_2\bsm{n}+\bsm{n}_1
\end{equation}
where $\bsm{n}$ and $\bsm{n}_1$ are $\mathcal{CCN}(0,\sigma^2\bsm{I})$. The complex channel coefficients $h_1$, $h_2$, $g_1$ and $g_2$ are assumed to remain fixed during the transmission of the $L$ pilot symbols and $N$ data symbols. 

We assume that both terminals employ square QAM modulation with possibly different modulation orders and transmission powers. Without loss of generality, we focus on the derivation of the CRB for channel estimation at terminal $\mathcal{T}_1$. For square QAM modulation, the total number of constellation points is $M=2^{2p}$, where $p=1,2,3,\hd$. Denoting by $d$ the intersymbol distance and letting $d_p\triangleq\frac{d}{2}$, the set of constellation points used by $\mathcal{T}_2$ is given by $S=\{\pm d_p(2i-1)\pm\jmath d_p(2\ell-1)\}, i,\ell=1,\hdots,2^{p-1}$~\cite{CRB_NDA2010}. The average transmitted power at $\mathcal{T}_2$ is $P_2=\Exp{|s_{2k}|^2}=\frac{M-1}{6}d^2$. We also let $P_1$ be the avarage transmitted power at $\mathcal{T}_1$. Furthermore, we assume that the noise variance $\sigma^2$ is known at $\mathcal{T}_1$. In our work, we are interested in deriving the CRBs for the estimation of the composite channel parameters $a\triangleq h_1h_2$ and $b\triangleq g_1h_2$, which are sufficient for data detection.

\section{Cramer-Rao Bounds}
\label{CRB_section}
In this section, we derive the CRBs for the estimation of $a$ and $b$. In addition to deriving the exact CRB, we will also consider the modified CRB (MCRB)~\cite{gini2000} which is commonly used in the presence of random nuisance parameters and is more tractable than the exact CRB. In deriving these bounds, we also take into account the parameter $|h_2|^2$ which, though not required for detection, appears in the likelihood function and thus affects the estimation performance. The unknowns parameters are thus $a$, $b$ and $|h_2|^2$ and are collected into the real vector $\bsm{\theta}\triangleq[\Re\{a\},\Im\{a\}, \Re\{b\},\Im\{b\},|h_2|^2]^{T}$. 

\subsection{Exact Cramer-Rao Bound}
To derive the exact CRB for semi-blind channel estimation, we consider the joint likelihood of $\bsm{z}_t$ and $\bsm{z}$. Let $\tilde{\bsm{z}}\triangleq[\bsm{z}_t^{T},\bsm{z}^{T}]^{T}$, the joint likelihood is given by
\begin{equation}
\label{likelihood}
\begin{split}
f(\tilde{\bsm{z}};\bsm{\theta})&=\frac{1}{(\pi\sigma^2(A^2|h_2|^2+1))^{N+L}}e^{-\frac{\Vert\bsm{z}_t-Aa\bsm{t}_1-Ab\bsm{t}_2\Vert^2}{\sigma^2(A^2|h_2|^2+1)}}\\
&\hspace{7ex}\times\prod_{k=1}^{N}\frac{1}{M}\bigg(\sum\limits_{s_2\in S}e^{-\frac{|z_k-Aas_{1k}-Abs_2|^2}{\sigma^2(A^2|h_2|^2+1)}}\bigg).
\end{split}
\end{equation}
The corresponding log-likelihood function is
\begin{equation}
\label{loglikelihood}
\begin{split}
&\mathcal{L}(\tilde{\bsm{z}};\bsm{\theta})=-(N+L)\log(\pi C)-\frac{1}{C}\Vert\bsm{z}_p-Aa\bsm{t}_1-Ab\bsm{t}_2\Vert^2\\&-N\log M
+\sum\limits_{k=1}^{N}\log\bigg(\sum\limits_{s_2\in S}e^{-\frac{1}{C}|z_k-Aas_{1k}-Abs_2|^2}\bigg).
\end{split}
\end{equation}
where $C\triangleq\sigma^2(A^2|h_2|^2+1)$. Let $\bsm{I}({\theta})$ be the corresponding Fisher information matrix (FIM), and let $I_{\theta_i,\theta_j}$ be the joint Fisher information between the parameters $\theta_i$ and $\theta_j$, where $i,j=1,\hd,5$. Furthermore, to simplify our notation we let $a_{R}\triangleq\Re\{a\}$, $a_{I}\triangleq\Im\{a\}$, $b_{R}\triangleq\Re\{b\}$, $b_{I}\triangleq\Im\{b\}$ and $\tau\triangleq|h_2|^2$. The matrix $\bsm{I}({\theta})$ is given by
\begin{equation}
\label{submatrices}
\begin{split}
\bsm{I}({\theta})&=-\Exp{\frac{\partial^2 \mathcal{L}(\tilde{\bsm{z}};\bsm{\theta})}{\partial \bsm{\theta}\partial\bsm{\theta}^{T}}}\\
&=\begin{bmatrix}\bsm{I}_{aa}&\bsm{I}_{ab}&\bsm{I}_{a\tau}\\
\bsm{I}_{ab}^{T}&\bsm{I}_{bb}&\bsm{I}_{b\tau}\\
\bsm{I}_{a\tau}^{T}&\bsm{I}_{b\tau}^{T}&I_{
\tau,\tau}\end{bmatrix},
\end{split}
\end{equation}
where
\begin{equation}
\bsm{I}_{aa}=\begin{bmatrix}I_{a_R,a_R}&I_{a_R,a_I}\\
I_{a_{R},a_{I}}&I_{a_I,a_I}
\end{bmatrix},\ \bsm{I}_{ab}=\begin{bmatrix}I_{a_R,b_R}&I_{a_R,b_I}\\
I_{a_I,b_R}&I_{a_I,b_I}
\end{bmatrix},
\end{equation}
\begin{equation}
\bsm{I}_{bb}=\begin{bmatrix}I_{b_R,b_R}&I_{b_R,b_I}\\
I_{b_R,b_I}&I_{b_I,b_I}
\end{bmatrix},\  \bsm{I_{a\tau}}=\begin{bmatrix}I_{a_R,\tau}\\I_{a_I,\tau}
\end{bmatrix},\  \bsm{I_{b\tau}}=\begin{bmatrix}I_{b_R,\tau}\\I_{b_I,\tau}
\end{bmatrix}.
\end{equation}

Before proceeding to obtain closed-form expressions for the elements of $\bsm{I}(\theta)$, we will first factorize the likelihood function by taking into account the symmetric structure of square QAM modulation, following the approach proposed in~\cite{CRB_NDA2010}. This factorization will make it feasible to derive analytical expressions for the elements of $\bsm{I}(\theta)$. We begin by rewriting the likelihood function in~\eqref{likelihood} as 
\begin{equation}
\label{likelihood_MQAM}
\begin{split}
&f(\tilde{\bsm{z}};\bsm{\theta})=\frac{1}{(\pi C)^{N+L}M^{N}}e^{-\frac{1}{C}\Vert\bsm{z}_p-Aa\bsm{t}_1-Ab\bsm{t}_2\Vert^2}\times\\
&\hs{0.8}\prod_{k=1}^{N}\hs{0.4}\bigg(e^{-\frac{1}{C}|z_k-Aas_{1k}|^2}\hs{0.8}\sum\limits_{s_2\in S}\hs{0.8}e^{-\frac{A^2}{C}|b|^2|s_2|^2+\frac{2A}{C}\Re\{(z_k-Aas_{1k})^{*}bs_2\}}\hs{0.4}\bigg).
\end{split}
\end{equation}
Now, we let
\begin{equation}
\label{Dk_expression0}
D_k(\bsm{\theta})\triangleq\sum\limits_{s_2\in S}e^{-\frac{1}{C}A^2|b|^2|s_2|^2+\frac{2A}{C}\Re\{(z_k-Aas_{1k})^{*}bs_2\}}.
\end{equation}
The inherent symmetry of the constellation set allows us to write~\eqref{Dk_expression0} as a sum over the symbols in the first quadrant. If $Q_1$ is the set of constellation symbols that lie in the first quadrant, $S$ can be partitioned as $S=Q_1\cup(-Q_1)\cup Q_1^{*}\cup(-Q_1^{*})$. Hence, we may rewrite $D_k(\bsm{\theta})$ as:
\begin{equation}
\label{Dk_expression}
\begin{split}
D_k(\bsm{\theta})=&\sum\limits_{s_2\in Q_1}e^{-\frac{1}{C}A^2|b|^2|s_2|^2}\bigg(e^{\frac{2A}{C}\Re\{(z_k-Aas_{1k})^{*}bs_2\}}\\
&+e^{\frac{2A}{C}\Re\{(z_k-Aas_{1k})^{*}bs_2^{*}\}}+e^{-\frac{2A}{C}\Re\{(z_k-Aas_{1k})^{*}bs_2\}}\\
&+e^{-\frac{2A}{C}\Re\{(z_k-Aas_{1k})^{*}bs_2^{*}\}}\bigg).
\end{split}
\end{equation}
Noting that $\Re\{(z_k-Aas_{1k})^{*}bs_2\}=\Re\{(z_k-Aas_{1k})^{*}b\}\Re\{s_2\}-\Im\{(z_k-Aas_{1k})^{*}b\}\Im\{s_2\}$, we rewrite~\eqref{Dk_expression} as
\begin{equation}
\label{Dk_expression2}
\begin{split}
D_k(\bsm{\theta})&=\hs{0.4}4\hs{1}\sum\limits_{s_2\in Q_1}\hs{1}e^{-\frac{1}{C}A^2|b|^2|s_2|^2}\\
&\times\cosh\left[\frac{2A}{C}\Re\{(z_k-Aas_{1k})^{*}b\}\Re\{s_2\}\right]\\
&\times\cosh\left[\frac{2A}{C}\Im\{(z_k-Aas_{1k})^{*}b\}\Im\{s_2\}\right].
\end{split}
\end{equation}
Moreover, for $s_2\in Q_1$, we have that $\Re\{s_2\},\Im\{s_2\}\in\{(2i-1)d_p\},\ i=1,\hdots,2^{p-1}$. Hence,~\eqref{Dk_expression2} becomes
\begin{equation}
\label{Dk_expression3}
\begin{split}
D_k(\bsm{\theta})&=4\sum\limits_{i=1}^{2^{p-1}}\sum\limits_{\ell=1}^{2^{p-1}}e^{-\frac{1}{C}A^2|b|^2((2i-1)^2+(2\ell-1)^2)d_p^2}\times\\
&\cosh\left[\frac{2A}{C}(2i-1)d_p\Re\{(z_k-Aas_{1k})^{*}b\}\right]\times\\
&\cosh\left[\frac{2A}{C}(2\ell-1)d_p\Im\{(z_k-Aas_{1k})^{*}b\}\right].
\end{split}
\end{equation}
From~\eqref{Dk_expression3}, we can see that, similar to the case in~\cite{CRB_NDA2010}, $D_k(\bsm{\theta})$ is the product of two terms, one depending on $\Re\{(z_k-Aas_{1k})^{*}b\}$ and the other depending on $\Im\{(z_k-Aas_{1k})^{*}b\}$:
\begin{equation}
\label{Dk_expression4}
D_k(\bsm{\theta})=4F_{\theta}(u_k)F_{\theta}(v_k),
\end{equation}
where 
\begin{equation}
\label{Ftheta_def}
F_{\theta}(t)\triangleq\sum\limits_{i=1}^{2^{p-1}}e^{-\frac{1}{C}A^2d_p^2|b|^2(2i-1)^2}\cosh\left(\frac{2Ad_p}{C}(2i-1)t\right),
\end{equation}
\begin{equation}
\label{Ui_def}
\begin{split}
u_k&\triangleq\Re\{(z_k-Aas_{1k})^{*}b\}\\
&=\Re\{z_k-Aas_{1k}\}b_R+\Im\{z_k-Aas_{1k}\}b_I
\end{split}
\end{equation}
and
\begin{equation}
\label{Vi_def}
\begin{split}
v_k&\triangleq\Im\{(z_k-Aas_{1k})^{*}b\}\\
&=\Re\{z_k-Aas_{1k}\}b_I-\Im\{z_k-Aas_{1k}\}b_R.
\end{split}
\end{equation}
Before proceeding, we simplify our notation by letting $\beta_i\triangleq\frac{Ad_p}{C}(2i-1)$ and $\gamma_i\triangleq\frac{A^2d_p^2}{C}(2i-1)^2$ for $i=1,\hdots,2^{p-1}$. Using the newly defined $\beta_i$, $\gamma_i$, we can write $F_{\theta}(t)$ as
\begin{equation}
F_{\theta}(t)\triangleq\sum\limits_{i=1}^{2^{p-1}}e^{-\gamma_i|b|^2}\cosh\left(2\beta_it\right).
\end{equation}
Using~\eqref{Dk_expression4}, the likelihood function becomes
\begin{equation}
\label{likelihood_MQAM2}
\begin{split}
f(\tilde{\bsm{z}};\theta)&=\frac{1}{(\pi C)^{N+L}}e^{-\frac{1}{C}\Vert\bsm{z}_p-Aa\bsm{t}_1-Ab\bsm{t}_2\Vert^2}\\
&\times\prod_{k=1}^{N}\frac{1}{M}\bigg(e^{-\frac{1}{C}|z_k-Aas_{1k}|^2}4F_{\theta}(u_k)F_{\theta}(v_k)\bigg).
\end{split}
\end{equation}
As we shall see shortly, the RVs $u_k$ and $v_k$ are independent, a fact which will simplify our derivation of the FIM. It is also useful to define two new RVs, $x_k\triangleq\Re\{z_k-Aas_{1k}\}$ and $y_k\triangleq\Im\{z_k-Aas_{1k}\}$. The pairs $\{u_k,v_k\}$ and pair $\{x_k,y_k\}$ are related through the following linear transformation
\begin{equation}
\label{linear_transformation}
\begin{bmatrix}u_k\\v_k\end{bmatrix}=\begin{bmatrix}b_R&b_I\\b_I&-b_R\end{bmatrix}\begin{bmatrix}x_k\\y_k\end{bmatrix}.
\end{equation}
Both pairs of RVs will be used in deriving the elements of the $\bsm{I}(\theta)$. It is easy to see that the joint PDF of $x_k$ and $y_k$ is  
\begin{equation}
\label{PDF_XY}
f_{X,Y}(x_k,y_k)\hs{0.5}=\hs{0.5}\frac{4}{\pi MC}e^{-\frac{x_k^2+y_k^2}{C}}\hs{0.4}F_{\theta}(b_Rx_k+b_Iy_k)F_{\theta}(b_Ix_k-b_Ry_k).
\end{equation}
Using~\eqref{linear_transformation} and~\eqref{PDF_XY}, we obtain the joint PDF of $u_k$ and $v_k$:
\begin{equation}
\label{PDF_UV}
f_{U,V}(u_k,v_k)=\frac{4}{\pi MC|b|^2}e^{-\frac{u_k^2+v_k^2}{C|b|^2}}F_{\theta}(u_k)F_{\theta}(v_k).
\end{equation}
It is clear from~\eqref{PDF_UV} that the RVs $u_k$ and $v_k$ are independent and identically distributed (i.i.d.) with respective PDFs
\begin{equation}
\label{PDF_Ui}
f_U(u_k)=\frac{2}{\sqrt{M\pi C|b|^2}}e^{-\frac{u_k^2}{C|b|^2}}F_{\theta}(u_k),
\end{equation}
and
\begin{equation}
\label{PDF_Vi}
f_V(v_k)=\frac{2}{\sqrt{M\pi C|b|^2}}e^{-\frac{v_k^2}{C|b|^2}}F_{\theta}(v_k).
\end{equation}

Going back to the log-likelihood function in~\eqref{loglikelihood}, we may now rewrite it as
\begin{equation}
\label{loglikelihood_QAM}
\begin{split}
\mathcal{L}(\tilde{\bsm{z}};\bsm{\theta})=&-(N+L)\log(\pi C)-\frac{1}{C}\Vert\bsm{z}_p-Aa\bsm{t}_1-Ab\bsm{t}_2\Vert^2+\\
&N\log\frac{4}{M}-\frac{1}{C}\Vert \bsm{z}-Aa\bsm{s}_1\Vert^2+\\
&\sum\limits_{k=1}^{N}\log F_{\theta}(u_k)+\sum\limits_{k=1}^{N}\log F_{\theta}(v_k).
\end{split}
\end{equation}
From~\eqref{loglikelihood_QAM} we see that the main task in obtaining analytical expressions for the elements of $\bsm{I}(\theta)$ is the evaluation of the expectations $\Exp{\frac{\partial^2\log F_{\theta}(u_k)}{\partial \theta_i\partial\theta_j}}$ and $\Exp{\frac{\partial^2\log F_{\theta}(v_k)}{\partial \theta_i\partial\theta_j}}$ for $i,j=1,\hd,5$. Letting
\begin{equation}
B_k^{(ij)}\triangleq\frac{\frac{\partial^2F_{\theta}(u_k)}{\partial \theta_i\partial\theta_j}}{F_{\theta}(u_k)},\ \ \ \ \ G_k^{(ij)}\triangleq\frac{\frac{\partial F_{\theta}(u_k)}{\partial \theta_i}\frac{\partial F_{\theta}(u_k)}{\partial \theta_j}}{F_{\theta}(u_k)^2},
\end{equation}
and
\begin{equation}
H_k^{(ij)}\triangleq\frac{\frac{\partial^2F_{\theta}(v_k)}{\partial \theta_i\partial\theta_j}}{F_{\theta}(v_k)},\ \ \ \ \ W_k^{(ij)}\triangleq\frac{\frac{\partial F_{\theta}(v_k)}{\partial \theta_i}\frac{\partial F_{\theta}(v_k)}{\partial \theta_j}}{F_{\theta}(v_k)^2},
\end{equation}
we have that
\begin{equation}
\frac{\partial^2\log F_{\theta}(u_k)}{\partial \theta_i\partial\theta_j}=B_k^{(ij)}-G_k^{(ij)},
\end{equation}
and
\begin{equation}
\frac{\partial^2\log F_{\theta}(v_k)}{\partial \theta_i\partial\theta_j}=H_k^{(ij)}-W_k^{(ij)}.
\end{equation}

Despite the factorization of the log-likelihood function, the derivation of analytical expressions for the elements of $\bsm{I}({\theta})$ requires tedious calculations. Due to space limitations, we will provide detailed derivations for only some of these elements. For the remaining elements we will provide only the resulting analytical expressions. We begin with the first diagonal element of the FIM, $I_{a_R,a_R}$. We have
\begin{equation}
\label{exp_RaRa1}
\begin{split}
&\Exp{\frac{\partial^2\mathcal{L}(\tilde{\bsm{z}};\bsm{\theta})}{\partial a_R^2}}=-\frac{2A^2}{C}\bsm{t}_1^{H}\bsm{t}_1-\frac{2A^2}{C}\bsm{s}_1^{H}\bsm{s}_1\\
&+\sum\limits_{k=1}^{N}\Exp{B_k^{(11)}-G_k^{(11)}}+\sum\limits_{k=1}^{N}\Exp{H_k^{(11)}-W_k^{(11)}}.\\
\end{split}
\end{equation}
We show in Appendix~\ref{appendixA} that
\begin{equation}
\label{Term11_1}
\Exp{B_k^{(11)}}=\frac{8A^2}{\sqrt{M}}\sum\limits_{i=1}^{2^{p-1}}\beta_i^2\Re\{s_{1k}^{*}b\}^2,
\end{equation}
and
\begin{equation}
\label{Term11_2}
\Exp{H_k^{(11)}}=\frac{8A^2}{\sqrt{M}}\sum\limits_{i=1}^{2^{p-1}}\beta_i^2\Im\{s_{1k}^{*}b\}^2.
\end{equation}
To obtain $\Exp{G_k^{(11)}}$, we need the first derivative of $F_{\theta}(u_k)$ with respect to $a_R$, which is given by
\begin{equation}
\label{appA1}
\begin{split}
\frac{\partial F_{\theta}(u_k)}{\partial a_R}=-2A\Re\{s_{1k}^{*}b\}\sum\limits_{i=1}^{2^{p-1}}\beta_ie^{-\gamma_i|b|^2}
\sinh[2\beta_i u_k].
\end{split}
\end{equation}
From~\eqref{appA1}, we see that we can evaluate $\Exp{G_k^{(11)}}$ using the PDF $f_U(u_k)$ in~\eqref{PDF_Ui}. We thus obtain
\begin{equation}
\label{Term11_3}
\Exp{G_k^{(11)}}=\Re\{s_{1k}^{*}b\}^2\Gamma_1,
\end{equation}
where
\begin{equation}
\Gamma_1=\frac{8A^2}{\sqrt{\pi MC|b|^2}}\int_{-\infty}^{\infty}\frac{f^2(t)}{F_{\theta}(t)}e^{-\frac{t^2}{C|b|^2}}dt,
\end{equation}
and
\begin{equation}
\label{def_ft}
f(t)=\sum\limits_{i=1}^{2^{p-1}}\beta_ie^{-\gamma_i|b|^2}\sinh[2\beta_i t].
\end{equation}
Moreover, it can be easily verified that
\begin{equation}
\label{Term11_4}
\Exp{W_k^{(11)}}=\Im\{s_{1k}^{*}b\}^2\Gamma_1,
\end{equation}
Hence,
\begin{equation}
\label{exp_RaRa2}
\begin{split}
&I_{a_R,a_R}=\frac{2A^2}{C}\bsm{t}_1^{H}\bsm{t}_1+\frac{2A^2}{C}\sum\limits_{k=1}^{N}\bsm{s}_1^{H}\bsm{s}_1\\
&-|b|^2\bsm{s}_1^{H}\bsm{s}_1\left(\frac{8A^2}{\sqrt{M}}\sum\limits_{i=1}^{2^{p-1}}\beta_i^2-\Gamma_1\right).
\end{split}
\end{equation}

Regarding the second diagonal element of $\bsm{I}(\theta)$, it can be easily shown that $I_{a_I,a_I}=I_{a_R,a_R}$.

For the third diagonal element of $\bsm{I}(\theta)$, we have
\begin{equation}
\label{exp_RbRb}
\begin{split}
\Exp{\frac{\partial^2\mathcal{L}(\tilde{\bsm{z}};\bsm{\theta})}{\partial b_R^2}}&=-\frac{2A^2}{C}\bsm{t}_2^{H}\bsm{t}_2
+\sum\limits_{k=1}^{N}\Exp{B_k^{(33)}-G_k^{(33)}}\\
&+\sum\limits_{k=1}^{N}\Exp{H_k^{(33)}-W_k^{(33)}}.
\end{split}
\end{equation}
Furthermore, we show in Appendix~\ref{appendixB} that
\begin{equation}
\begin{split}
\label{Term33_1}
\Exp{B_k^{(33)}}=\Exp{H_k^{(33)}}=\frac{16}{M}\sum\limits_{i=1}^{2^{p-1}}\sum\limits_{\ell=1}^{2^{p-1}}\gamma_i\gamma_{\ell}b_I^2.
\end{split}
\end{equation}

To find $\Exp{G_k^{(33)}}$, we obtain the derivative of $F_{\theta}(u_k)$ with respect to $b_R$, which is given by
\begin{equation}
\label{dF1Rb}
\begin{split}
&q_1(x_k,y_k)\triangleq\frac{\partial F_{\theta}(b_Rx_k+b_Iy_k)}{\partial b_R}\\
&=-\sum\limits_{i=1}^{2^{p-1}}2\gamma_i b_Re^{-\gamma_i|b|^2}\cosh[2\beta_i(b_Rx_k+b_I y_k)]\\
&+\sum\limits_{i=1}^{2^{p-1}}2\beta_ix_ke^{-\gamma_i|b|^2}\sinh[2\beta_i(b_Rx_k+b_I y_k)].
\end{split}
\end{equation}
Let $\Gamma_2\triangleq\Exp{G_k^{(33)}}=\Exp{\frac{q_1(x_k,y_k)^2}{F_{\theta}(u_k)}}$. Clearly, $\Gamma_2$ cannot be evaluated using the PDF $f_U(u)$, and we need to use the joint PDF $f_{XY}(x,y)$ in~\eqref{PDF_XY}, which means that double integration is required. Using~\eqref{PDF_XY}, we obtain
\begin{equation}
\begin{split}
\Gamma_2=\frac{4}{\pi MC}\iint\limits_{-\infty}^{\hs{-3}\infty}\frac{q_1^2(x,y)F_{\theta}(b_Ix-b_Ry)}{F_{\theta}(b_Rx+b_Iy)}e^{-\frac{x^2+y^2}{C}}dxdy.
\end{split}
\end{equation}
Moreover, it can be easily verified that $\Exp{W_k^{(33)}}=\Exp{G_k^{(33)}}$, which implies that
\begin{equation}
\label{exp_RbRb2}
\begin{split}
I_{b_R,b_R}=&\frac{2A^2}{C}\bsm{t}_2^{H}\bsm{t}_2-\frac{32N}{M}\sum\limits_{i=1}^{2^{p-1}}\sum\limits_{\ell=1}^{2^{p-1}}\gamma_i\gamma_{\ell}b_I^2\\
&+2N\Gamma_2.
\end{split}
\end{equation}

A very similar approach can be followed to evaluate $I_{b_I,b_I}$, thus obtaining
\begin{equation}
\label{exp_IbIb}
\begin{split}
I_{b_I,b_I}=&\frac{2A^2}{C}\bsm{t}_2^{H}\bsm{t}_2-\frac{32N}{M}\sum\limits_{i=1}^{2^{p-1}}\sum\limits_{\ell=1}^{2^{p-1}}\gamma_i\gamma_{\ell}b_R^2+2N\Gamma_3,
\end{split}
\end{equation}
where
\begin{equation}
\Gamma_3=\frac{4}{\pi MC}\iint\limits_{-\infty}^{\hs{-3}\infty}\frac{q_2^2(x,y)F_{\theta}(b_Ix-b_Ry)}{F_{\theta}(b_Rx+b_Iy)}e^{-\frac{x^2+y^2}{C}}dxdy
\end{equation}
and
\begin{equation}
\label{dF1Ib}
\begin{split}
&q_2(x_k,y_k)\triangleq\frac{\partial F_{\theta}(u_k)}{\partial b_I}\\
&=-\sum\limits_{i=1}^{2^{p-1}}2\gamma_i b_Ie^{-\gamma_i|b|^2}\cosh[2\beta_i(b_Rx_k+b_I y_k)]\\
&+\sum\limits_{i=1}^{2^{p-1}}2\beta_iye^{-\gamma_i|b|^2}\sinh[2\beta_i(b_Rx_k+b_I y_k)].
\end{split}
\end{equation}

For the fifth diagonal element of $\bsm{I}(\theta)$, $I_{\tau,\tau}$, we have
\begin{equation}
\label{exp_tautau}
\begin{split}
&\Exp{\frac{\partial^2\mathcal{L}(\tilde{\bsm{z}};\bsm{\theta})}{\partial \tau^2}}=(N+L)\frac{A^4\sigma^4}{C^2}\\
&-\frac{2A^4\sigma^4}{C^3}\left(\Exp{\Vert\bsm{z}_p-Aa\bsm{t}_1-Ab\bsm{t}_2\Vert^2}
+\Exp{\Vert\bsm{z}-Aa\bsm{s}_1\Vert^2}\right)\\
&+\sum\limits_{k=1}^{N}\Exp{B_k^{(55)}-G_k^{(55)}}+\sum\limits_{k=1}^{N}\Exp{H_k^{(55)}-W_k^{(55)}}.
\end{split}
\end{equation}
It can be easily shown that $\Exp{\Vert\bsm{z}_p-Aa\bsm{t}_1-Ab\bsm{t}_2\Vert^2}=LC$ and $\Exp{\Vert\bsm{z}-Aa\bsm{s}_1\Vert^2}=NA^2|b|^2P_2+NC$. Moreover, we show in Appendix~\ref{appendixC} that 
\begin{equation}
\label{EqBk55}
\Exp{B_k^{(55)}}=\Exp{H_k^{(55)}}=\sum\limits_{i=1}^{2^{p-1}}\frac{2A^4\sigma^4}{\sqrt{M}}\left(\hs{0.25}\beta_i^4|b|^4+\frac{4}{C}\beta_i^2|b|^2\hs{0.25}\right)\hs{0.5}.
\end{equation}

To obtain $\Exp{G_k^{(55)}}$, we first take the derivative $F_{\theta}(u_k)$ with respect to $\tau$: 
\begin{equation}
\begin{split}
q_3(u_k)&\triangleq\frac{\partial F_{\theta}(u_k)}{\partial\tau}=\sum\limits_{i=1}^{2^{p-1}}A^2\sigma^2\beta_i^2|b|^2e^{-\gamma_i|b|^2}\cosh[2\beta_iu_k]\\
&-\sum\limits_{i=1}^{2^{p-1}}\frac{2A^2\sigma^2}{C}\beta_iu_ke^{-\gamma_i|b|^2}\sinh[2\beta_iu_k].
\end{split}
\end{equation}
Letting $\Gamma_4\triangleq\Exp{G_k^{(55)}}$, we thus get
\begin{equation}
\Gamma_4=\frac{2}{\sqrt{\pi MC|b|^2}}\int\limits_{-\infty}^{\hs{-3}\infty}\frac{q_3(t)^2}{F_{\theta}(t)}e^{-\frac{t^2}{C|b|^2}}dt.
\end{equation}
Clearly, we also have that $\Exp{W_k^{55}}=\Gamma_4$. Using the above results, we get
\begin{equation}
\label{exp_tautau2}
\begin{split}
&I_{\tau,\tau}=-(N+L)\frac{A^4\sigma^4}{C^2}\\
&+\frac{2A^4\sigma^4}{C^3}\left((N+L)C+NA^2|b|^2P_2\right)\\
&-4N\sum\limits_{i=1}^{2^{p-1}}\frac{2A^4\sigma^4}{\sqrt{M}}\left(\beta_i^4|b|^4+\frac{4}{C}\beta_i^2|b|^2\right)+2N\Gamma_4.
\end{split}
\end{equation}

We now consider the off-diagonal elements of $\bsm{I}(\theta)$. First, we show in Appendix~\ref{appendixD} that $I_{a_R,a_I}=0$. The remaining elements of $\bsm{I}(\theta)$ can be obtained using similar approaches to those employed so far, and we will only provide the resulting analytical expressions. Going back to the sub-block matrices of $\bsm{I}(\theta)$ in~\eqref{submatrices}, the matrices $\bsm{I}_{aa}$ and $\bsm{I}_{bb}$ are given in~\eqref{eqnIa2} and~\eqref{eqnIb2}, respectively, at the top of the next page, where
\begin{figure*}[!t]
\normalsize
\begin{equation}
\label{eqnIa2}
\bsm{I}_{aa}=\begin{bmatrix}\frac{2A^2}{C}\bsm{t}_1^{H}\bsm{t}_1+\frac{2A^2}{C}\sum\limits_{k=1}^{N}\bsm{s}_1^{H}\bsm{s}_1
-|b|^2\bsm{s}_1^{H}\bsm{s}_1\left(\hs{0.5}\frac{8A^2}{\sqrt{M}}\sum\limits_{i=1}^{2^{p-1}}\beta_i^2-\Gamma_1\hs{0.5}\right)&\hs{3}0\\
0&\hs{3}\frac{2A^2}{C}\bsm{t}_1^{H}\bsm{t}_1+\frac{2A^2}{C}\sum\limits_{k=1}^{N}\bsm{s}_1^{H}\bsm{s}_1
-|b|^2\bsm{s}_1^{H}\bsm{s}_1\left(\hs{0.5}\frac{8A^2}{\sqrt{M}}\sum\limits_{i=1}^{2^{p-1}}\beta_i^2-\Gamma_1\hs{0.5}\right)
\end{bmatrix}
\end{equation}
\begin{equation}
\label{eqnIb2}
\bsm{I}_{bb}=\begin{bmatrix}\frac{2A^2}{C}\bsm{t}_2^{H}\bsm{t}_2-\frac{32N}{M}\sum\limits_{i=1}^{2^{p-1}}\sum\limits_{\ell=1}^{2^{p-1}}\gamma_i\gamma_{\ell}b_I^2+2N\Gamma_2&\frac{32N}{M}\sum\limits_{i=1}^{2^{p-1}}\sum\limits_{\ell=1}^{2^{p-1}}\gamma^2b_Rb_I(2i-1)^2(2\ell-1)^2
+2N\Gamma_5\\
\frac{32N}{M}\sum\limits_{i=1}^{2^{p-1}}\sum\limits_{\ell=1}^{2^{p-1}}\gamma^2b_Rb_I(2i-1)^2(2\ell-1)^2
+2N\Gamma_5&\frac{2A^2}{C}\bsm{t}_2^{H}\bsm{t}_2-\frac{32N}{M}\sum\limits_{i=1}^{2^{p-1}}\sum\limits_{\ell=1}^{2^{p-1}}\gamma_i\gamma_{\ell}b_R^2+2N\Gamma_3
\end{bmatrix}
\end{equation}
\hrulefill
\end{figure*}
\begin{equation}
\Gamma_5\hs{0.25}=\hs{0.25}\frac{4}{\pi MC}\hs{0.5}\iint\limits_{-\infty}^{\hs{-3}\infty}\frac{q_1(x,y)q_2(x,y)F_{\theta}(b_Ix-b_Ry)}{F_{\theta}(b_Rx+b_Iy)}e^{-\frac{x^2+y^2}{C}}dxdy.
\end{equation}
For the remaining submatrices, we have that
\begin{equation}
\label{eqnIab}
\bsm{I}_{ab}=\begin{bmatrix}\frac{2A^2}{C}\Re\{\bsm{t}_1^{H}\bsm{t}_2\}&-\frac{2A^2}{C}\Im\{\bsm{t}_1^{H}\bsm{t}_2\}\\
\frac{2A^2}{C}\Im\{\bsm{t}_1^{H}\bsm{t}_2\}&\frac{2A^2}{C}\Re\{\bsm{t}_1^{H}\bsm{t}_2\}
\end{bmatrix},
\end{equation}
\begin{equation}
\label{eqnIatau}
\bsm{I}_{a\tau}=\begin{bmatrix}0\\ 0\end{bmatrix}
\end{equation}
and
\begin{equation}
\label{eqnIbtau}
\bsm{I}_{b\tau}=\begin{bmatrix} -8N\sum\limits_{i=1}^{2^{p-1}}\frac{A^2\sigma^2}{\sqrt{M}}\beta_i^2b_R+2N\Gamma_6\\ -8N\sum\limits_{i=1}^{2^{p-1}}\frac{A^2\sigma^2}{\sqrt{M}}\beta_i^2b_I+2N\Gamma_7 \end{bmatrix}
\end{equation}
where
\begin{equation}
\begin{split}
\Gamma_6=\frac{4}{\pi MC}\iint\limits_{-\infty}^{\hs{-3}\infty}&\frac{q_3(b_Rx+b_Iy)q_1(x,y)}{F_{\theta}(b_Rx+b_Iy)}\\
&\times F_{\theta}(b_Ix-b_Ry)e^{-\frac{x^2+y^2}{C}} dxdy,
\end{split}
\end{equation} 
and
\begin{equation}
\begin{split}
\Gamma_7=\frac{4}{\pi MC}\iint\limits_{-\infty}^{\hs{-3}\infty}&\frac{q_3(b_Rx+b_Iy)q_2(x,y)}{F_{\theta}(b_Rx+b_Iy)}\\
&\times F_{\theta}(b_Ix-b_Ry)e^{-\frac{x^2+y^2}{C}} dxdy.
\end{split}
\end{equation}

Having derived the FIM matrix $\bsm{I}(\theta)$, the exact CRBs on $a$ and $b$ can be obtained by taking the inverse of $\bsm{I}(\theta)$. We have\footnote{The notation $[\bsm{A}]_{ij}$ is used to refer to the $(i,j)$th element of the matrix $\bsm{A}$.}
\begin{equation}
\label{CRB_a}
CRB_a=[\bsm{I}(\theta)^{-1}]_{11}+[\bsm{I}(\theta)^{-1}]_{22},
\end{equation}
and
\begin{equation}
\label{CRB_b}
CRB_b=[\bsm{I}(\theta)^{-1}]_{33}+[\bsm{I}(\theta)^{-1}]_{44}.
\end{equation}

The bounds in~\eqref{CRB_a} and~\eqref{CRB_b} provide convenient benchmarks for the performance of estimators of $a$ and $b$. For $N=0$, they cover the case of fully pilot-based estimation and for $L=0$ they cover the case of blind estimation\footnote{We note that in the absence of pilots the channel parameter $a$ remains identifiable thanks to the presence of known self-interference symbols. The parameter $b$, however, suffers from an inherent ambiguity and is only locally identifiable. The CRB is still defined in this case, as pointed out in~\cite{hero1990}. In practice, however, pilot symbols are required to resolve the ambiguity.}.  In Section~\ref{simulations}, we will use these bounds to compare the the semi-blind approach with the pilot-based and blind approaches. The obvious shortcoming of the exact CRBs is their high complexity. We next consider a more tractable alternative, the modified CRB (MCRB). 

\subsection{The Modified Cramer-Rao Bound}

Due to its tractability, the MCRB is commonly used in the presence of random nuisance parameters~\cite{gini2000}. In our case, the nuisance parameters are the data symbols $s_{2k}, k=1,\hd,N$. To obtain this bound, the data symbols are initially treated as deterministic unknowns, which greatly simplifies the likelihood function and the resulting FIM. The statistics of the data symbols are then taken into account by averaging the FIM over the data symbols. We call the resulting matrix the modified FIM (MFIM) and denote it by $\bsm{J}$. It can be shown that the matrix $\bsm{J}$ is given by~\eqref{MFIM} on top of the next page. Denoting by $MCRB_a$ and $MCRB_b$ the resulting bounds for parameters $a$ and $b$, respectively, it can be shown that 
\begin{figure*}[!t]
\normalsize
\begin{equation}
\label{MFIM}
\bsm{J}=\begin{bmatrix}\frac{2A^2}{C}\left(\bsm{t}_1^{H}\bsm{t}_1+\bsm{s}_1^{H}\bsm{s}_1\right)&0&\frac{2A^2}{C}\Re\{\bsm{t_1}^{H}\bsm{t}_2\}&-\frac{2A^2}{C}\Im\{\bsm{t_1}^{H}\bsm{t}_2\}&0\\
0&\frac{2A^2}{C}\left(\bsm{t}_1^{H}\bsm{t}_1+\bsm{s}_1^{H}\bsm{s}_1\right)&\frac{2A^2}{C}\left(\Im\{\bsm{t}_1^{H}\bsm{t}_2\}\right)&\frac{2A^2}{C}\Re\{\bsm{t_1}^{H}\bsm{t}_2\}&0\\
\frac{2A^2}{C}\Re\{\bsm{t_1}^{H}\bsm{t}_2\}&\frac{2A^2}{C}\left(\Im\{\bsm{t}_1^{H}\bsm{t}_2\}\right)&\frac{2A^2}{C}\left(\bsm{t}_2^{H}\bsm{t}_2+NP_2\right)&0&0\\
-\frac{2A^2}{C}\Im\{\bsm{t_1}^{H}\bsm{t}_2\}&\frac{2A^2}{C}\Re\{\bsm{t}_1^{H}\bsm{t}_2\}&0&\frac{2A^2}{C}\left(\bsm{t}_2^{H}\bsm{t}_2+NP_2\right)&0\\
0&0&0&0&(N+L)\frac{A^4\sigma^4}{C^2}
\end{bmatrix}
\end{equation}
\hrulefill
\end{figure*}
\begin{equation}
\begin{split}
MCRB_a=\frac{C\left(\bsm{t}_2^{H}\bsm{t}_2+NP_2\right)}{A^2\left((\bsm{t}_1^{H}\bsm{t}_1+\bsm{s}_1^{H}\bsm{s}_1)(\bsm{t}_2^{H}\bsm{t}_2+NP_2)-\bsm{t}_1^{H}\bsm{t}_2\bsm{t}_2^{H}\bsm{t}_1)\right)}
\end{split}
\end{equation}
and
\begin{equation}
\begin{split}
&MCRB_b=\frac{C}{A^2(\bsm{t}_2^{H}\bsm{t}_2+NP_2)}\\
&\times\left(1+\frac{\bsm{t}_1^{H}\bsm{t}_2\bsm{t}_2^{H}\bsm{t}_1}{\left((\bsm{t}_1^{H}\bsm{t}_1+\bsm{s}_1^{H}\bsm{s}_1)(\bsm{t}_2^{H}\bsm{t}_2+NP_2)-\bsm{t}_1^{H}\bsm{t}_2\bsm{t}_2^{H}\bsm{t}_1\right)}\right).
\end{split}
\end{equation}
The above expressions are much simpler and more tractable than those in~\eqref{CRB_a} and~\eqref{CRB_b}. However, as we shall see in Section~\ref{simulations}, they are loose and do not reflect the impact of the modulation order on the estimation accuracy. However, for low modulation orders they are close to the true bounds at high SNR. 

\section{Simulation Results}
\label{simulations}

In this section, we use MATLAB simulations to investigate the behavior of the derived CRBs for $a$ and $b$ for the cases of blind, semi-blind and pilot-based estimation. All plots are generated using the Monte-Carlo approach and are averaged over a set of $100$ independent realizations of the channel parameters $h_1$, $h_2$, $g_1$ and $g_2$. These realizations are generated by modelling $h_1$ and $h_2$ as correlated complex Gaussian random variables with mean zero, variance $1$, and a correlation coefficient $\varrho=0.3$. Similarly, we model $g_1$ and $g_2$ as correlated complex Gaussian random variables with the same mean, variance and correlation coefficient, but independent of $h_1$ and $h_2$. To generate correlated complex Gaussian random variables we follow the approach proposed in~\cite{tellambura_correlated}. In order to see the effect of the modulation order on the CRB, we consider four modulation orders, $M=4$, $M=16$, $M=64$ and $M=256$ in all our plots. In all our simulations, the pilots are generated using $M=4$, and the pilot vectors of the two terminals are are orthogonal to each other. The SNR is defined as $10\log\frac{P_2}{\sigma^2}$. 

\begin{figure}[htbp]
\centering
\includegraphics[width=3.7in, height=2.8in]{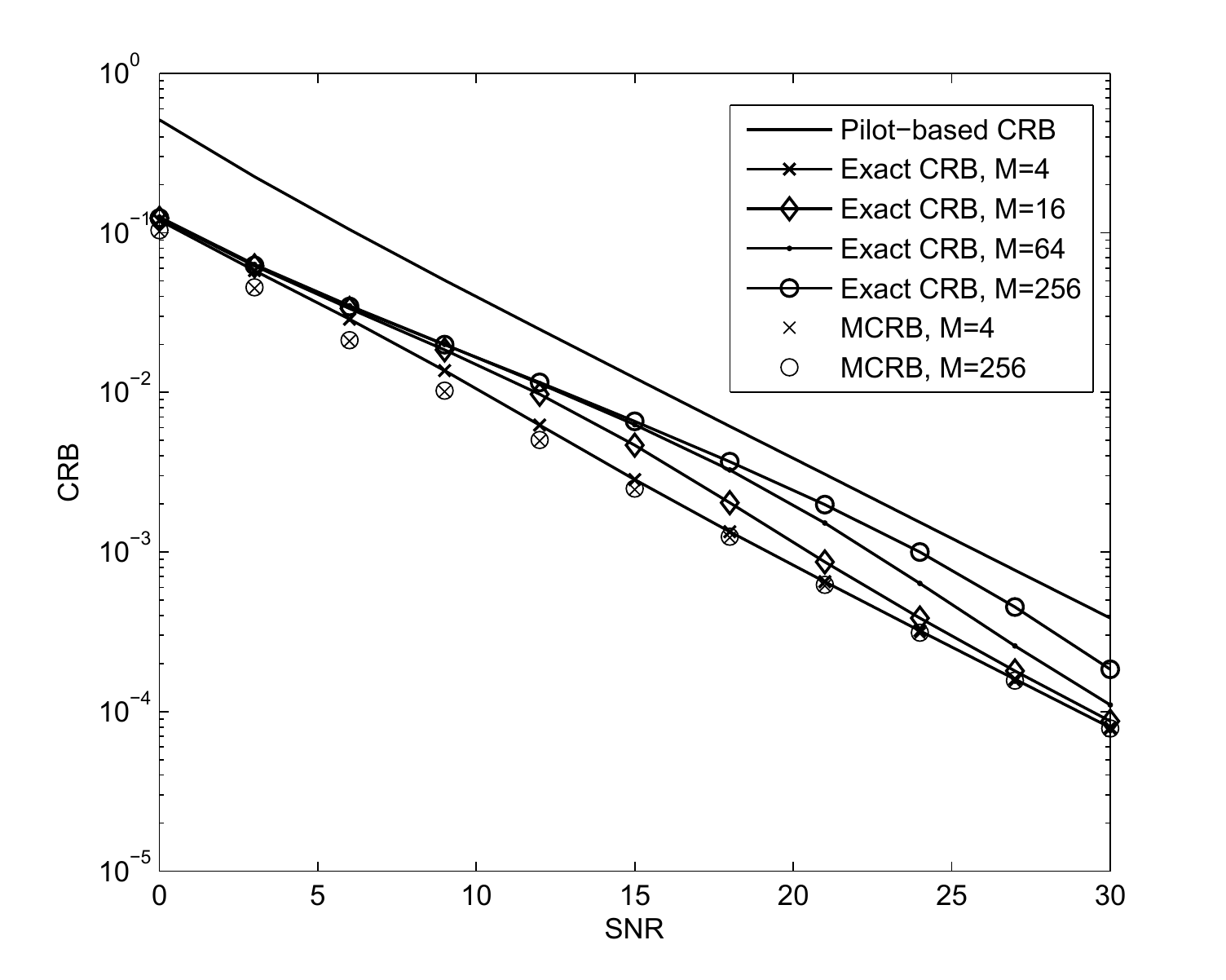}
\caption{Semi-blind and pilot-based CRBs for the estimation of $a$ plotted versus SNR for $N=32$ and $L=8$, and for $M=4,\ 16,\ 64,\ 256$. We also plot $MCRB_a$ for $M=4$ and $M=256$.}
\label{CRBa_SB_P}
\end{figure}

\begin{figure}[htbp]
\centering
\includegraphics[width=3.7in, height=2.8in]{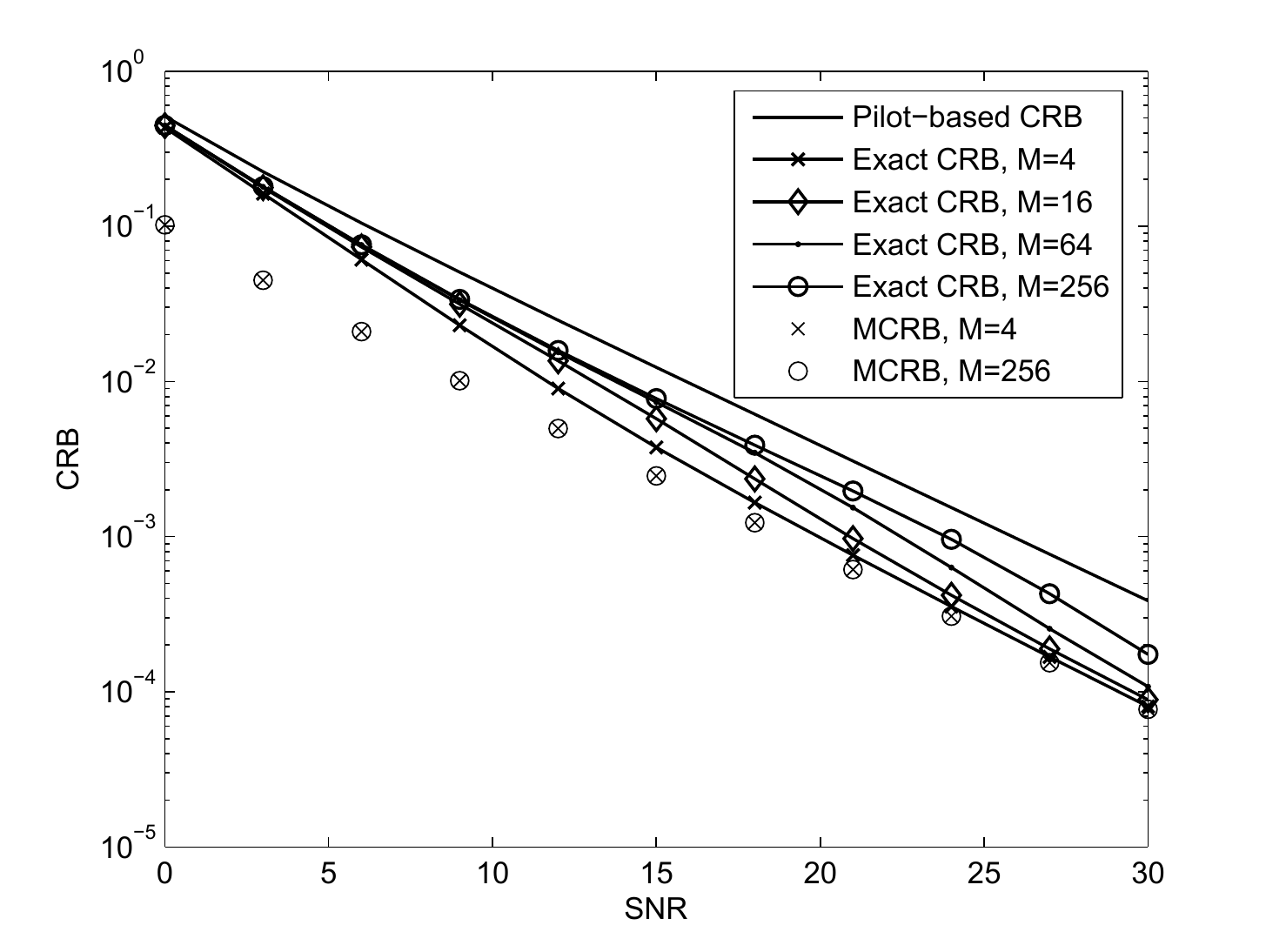}
\caption{Semi-blind and pilot-based CRBs for the estimation of $b$ plotted versus SNR for $N=32$ and $L=8$, and for $M=4,\ 16,\ 64,\ 256$. We also plot $MCRB_b$ for $M=4$ and $M=256$.}
\label{CRBb_SB_P}
\end{figure}

We begin by comparing the CRBs of the semi-blind approach and the pilot-based approach. In Figs.~\ref{CRBa_SB_P} and~\ref{CRBb_SB_P}, we plot versus SNR the semi-blind CRB and the pilot-based CRB for parameters $a$ and $b$, respectively. The number of pilots is $L=8$, and the number of transmitted data symbols is $N=32$. We also plot the corresponding MCRB for $M=4$ and $M=256$ in both figures. As we can see from both figures, the semi-blind CRB is substantially lower than the pilot-based bound for all 4 modulation orders. This shows that, by making use of the transmitted data symbols in addition to the pilots, the semi-blind approach can provide substantial gains in estimation accuracy over the pilot-based approach, thus providing a superior tradeoff between accuracy and spectral efficiency. We also see that the lower the modulation order the higher the achievable accuracy, with the best accuracy at $M=4$. In addition, Figs.~\ref{CRBa_SB_P} and~\ref{CRBb_SB_P} show that the MCRB is generally loose compared to the true CRB and is not sensitive to the modulation orders, and thus does not reflect the effect of the modulation order on the estimation accuracy. However, MCRB provides a good approximation of the true CRB at high SNR for $M=4$ and $M=16$.

\begin{figure}[htbp]
\centering
\includegraphics[width=3.7in, height=2.8in]{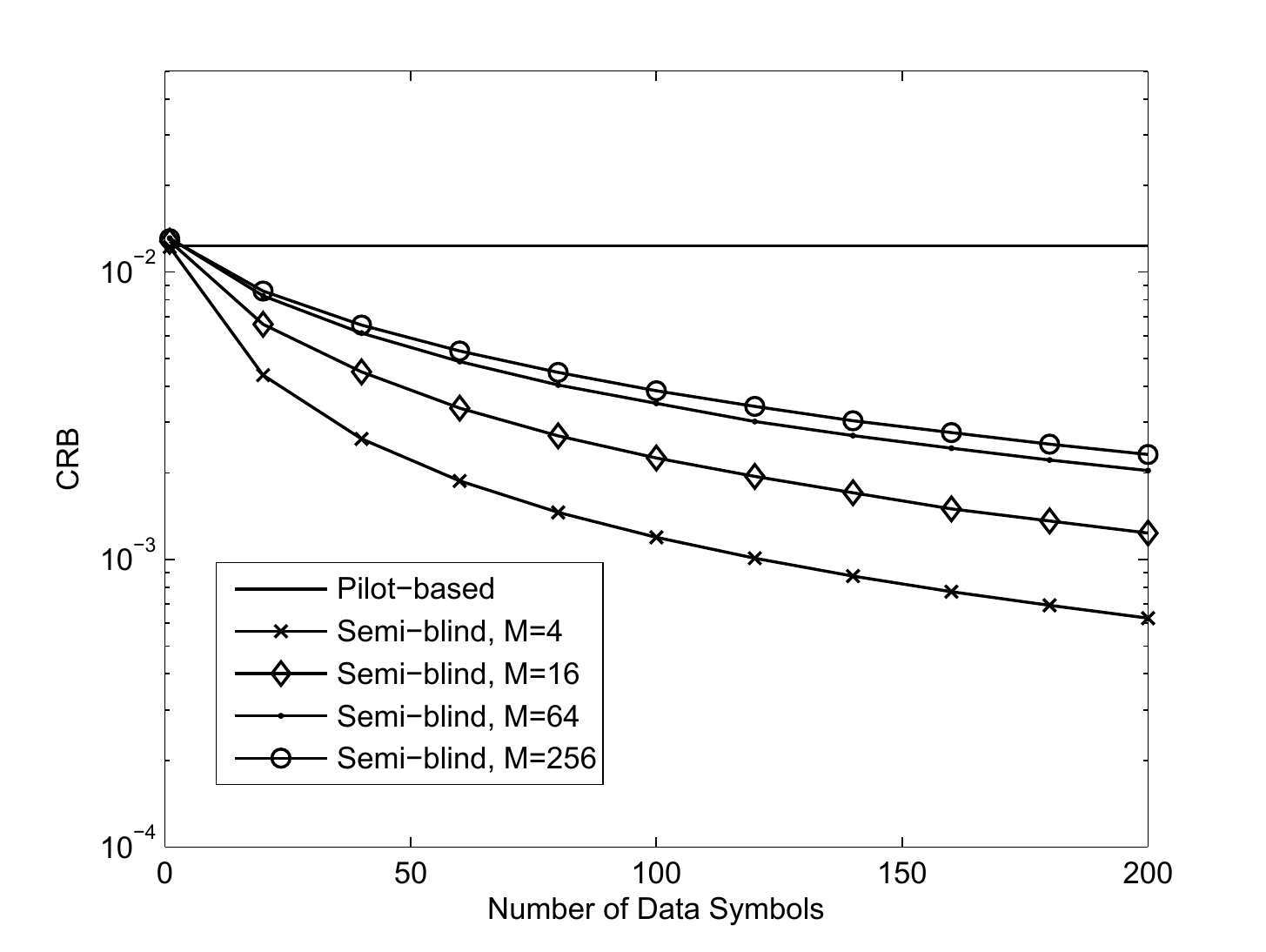}
\caption{Semi-blind CRB for the estimation of $a$ plotted versus $N$ for $L=8$, and for $M=4,\ 16,\ 64,\ 256$. The pilot-based CRB is shown as a reference.}
\label{CRBa_N}
\end{figure}

\begin{figure}[htbp]
\centering
\includegraphics[width=3.7in, height=2.8in]{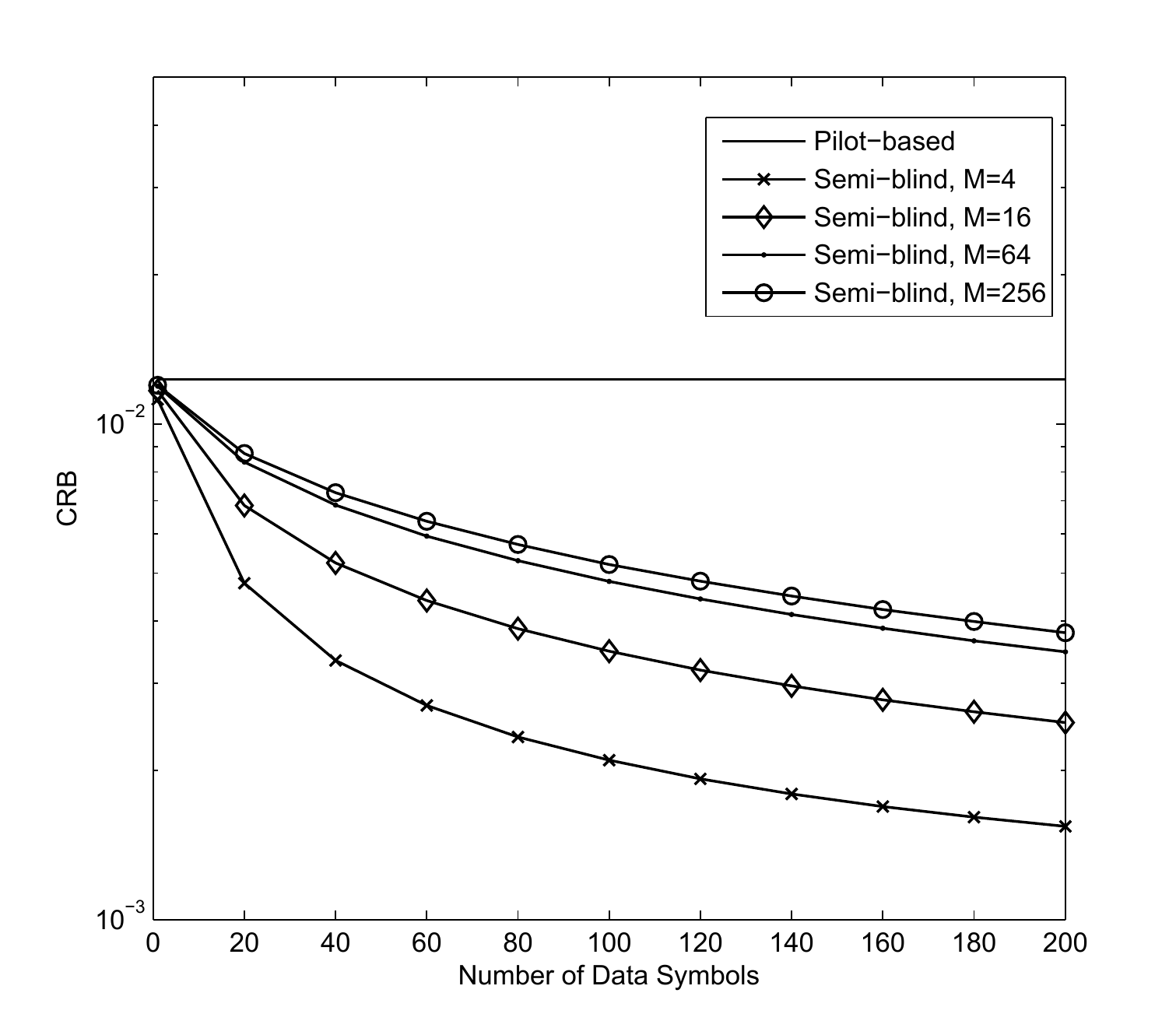}
\caption{Semi-blind CRB for the estimation of $b$ plotted versus $N$ for $L=8$, and for $M=4,\ 16,\ 64,\ 256$. The pilot-based CRB is shown as a reference.}
\label{CRBb_N}
\end{figure}

In Figs.~\ref{CRBa_N} and~\ref{CRBb_N}, we plot the semi-blind CRBs of $a$ and $b$ versus $N$, respectively. The number of pilots  is $L=8$, and the pilot-based CRB is plotted as a reference. As we see from both plots, as $N$ increases, the semi-blind approach can provide increasingly better accuracy compared to the pilot-based approach. The sample size $N$ is  constrained by the coherence time of the channel during which the channel parameters $a$, $b$ remain fixed. Hence, the longer the channel coherence time the more attractive the semi-blind approach becomes.

\begin{figure}[htbp]
\centering
\includegraphics[width=3.7in, height=2.8in]{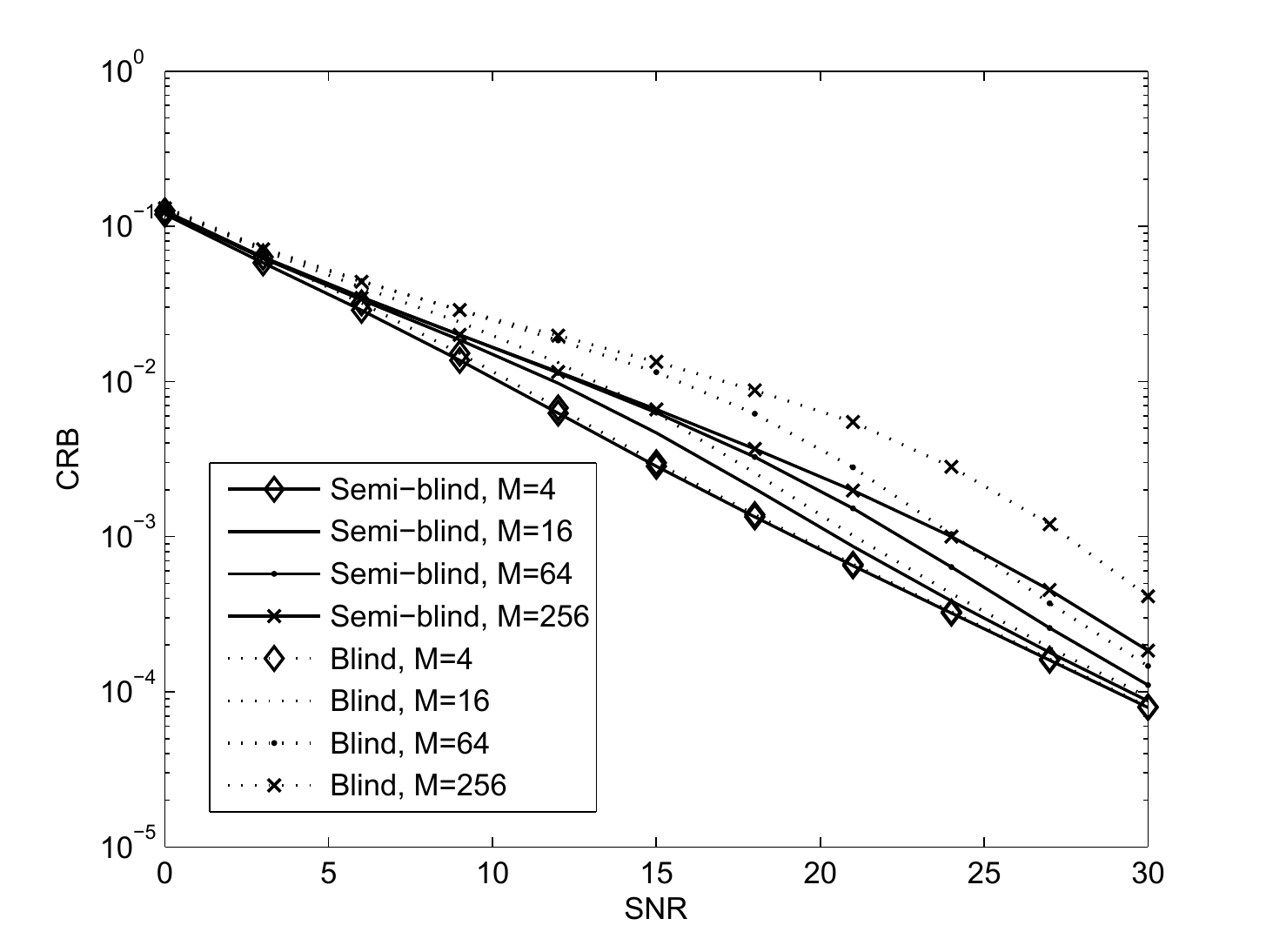}
\caption{Semi-blind and blind CRBs for the estimation of $a$ plotted versus SNR for $M=4,\ 16,\ 64,\ 256$. We use $N=32$ data symbols and $L=8$ pilots for the semi-blind case, and we use $N=40$ data symbols for the blind case.}
\label{CRBa_B}
\end{figure}

\begin{figure}[htbp]
\centering
\includegraphics[width=3.7in, height=2.8in]{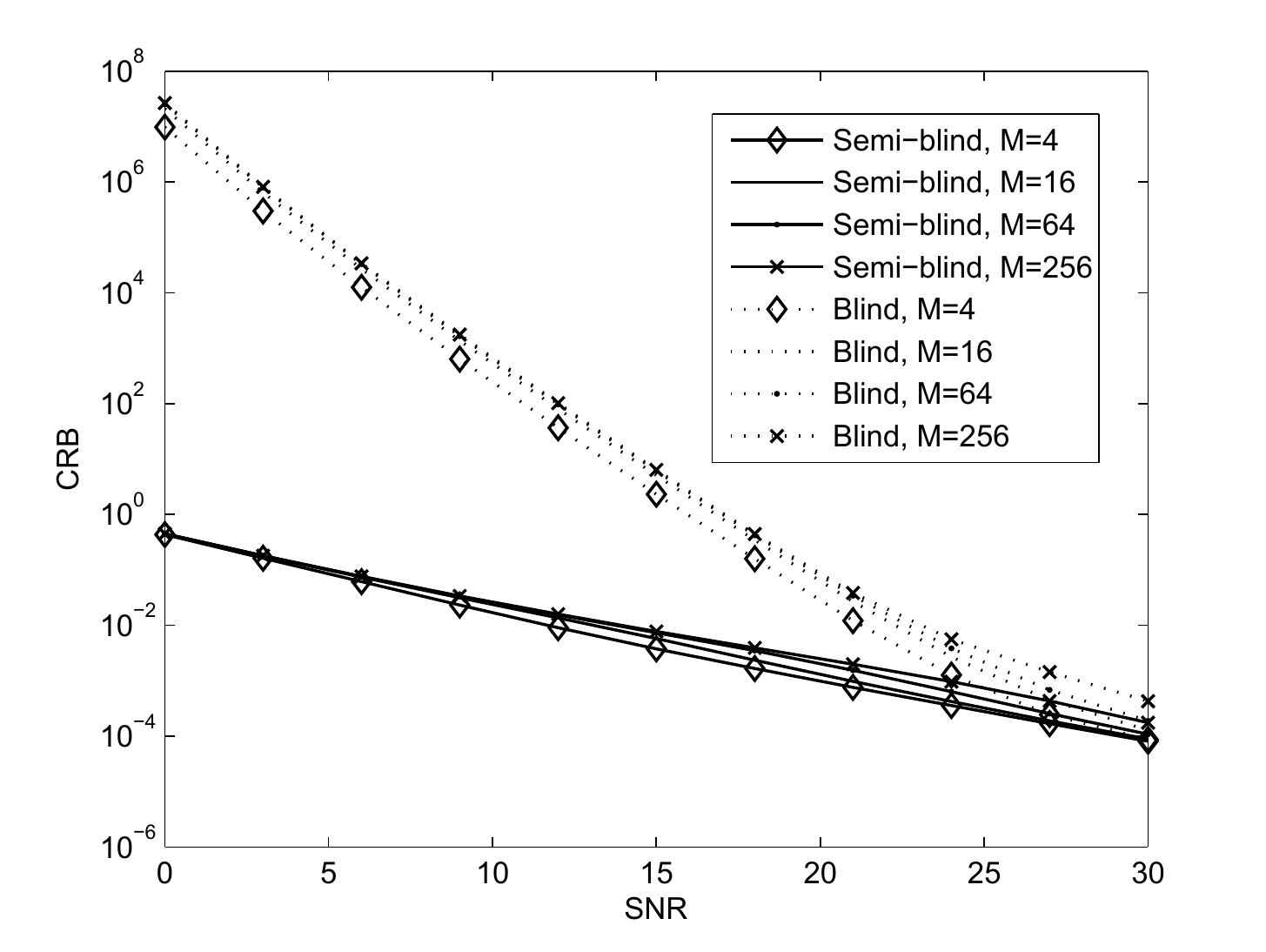}
\caption{Semi-blind and blind CRBs for the estimation of $b$ plotted versus SNR for $M=4,\ 16,\ 64,\ 256$. We use $N=32$ data symbols and $L=8$ pilots for the semi-blind case, and we use $N=40$ data symbols for the blind case.}
\label{CRBb_B}
\end{figure}

We next compare the CRBs of the semi-blind approach and the blind approach. In Figs.~\ref{CRBa_B} and~\ref{CRBb_B}, we plot versus SNR the semi-blind and blind CRBs for parameters $a$ and $b$, respectively. For the semi-blind case, $L=8$ pilots and $N=32$ data symbols are employed. For the blind case, $N=40$ data symbols are used. We see from Fig.~\ref{CRBa_B} that the semi-blind approach provides better accuracy for the estimation of $a$. As the modulation order increases, the gap in favor of the semi-blind approach becomes more significant. In Fig.~\ref{CRBb_B}, however, we see that the blind CRB for $b$ completely deteriorates at low SNR, reflecting the difficulty in the estimation of $b$ in the absence of pilots. The difference between the behaviors of the blind CRB for $a$ in Fig.~\ref{CRBa_B} and for $b$ in Fig.~\ref{CRBb_B} reflects the fact that in the absence of pilots the estimation of $a$ is an easier task than the estimation of $b$ due to the presence of the known self-interference symbols.

\section{Conclusions}
\label{conclusions}
In this paper, we derived the exact CRBs for semi-blind channel estimation in AF TWRNs employing square QAM modulation. As an more tractable alternative, we also derived the modified CRB. Using Monte-Carlo simulations, we showed that the semi-blind approach can provided substantial gains over the pilot-based approach. The semi-blind approach also seems to be more attractive than the blind approach, which deteriorates at low SNR. For future work, it remains to design efficient low-complexity semi-blind algorithms whose performance approaches the semi-blind CRB. 
 
\appendices

\section{}
\label{appendixA}
In this appendix, we prove~\eqref{Term11_1} and~\eqref{Term11_2}. Recalling that $u_k=\Re\{z_k-Aas_{1k}\}b_R+\Im\{z_k-Aas_{1k}\}b_I$, the second derivative of $F_{\theta}(u_k)$ with respect to $a_R$ is given by
\begin{equation}
\label{appA2}
\begin{split}
\frac{\partial^2 F_{\theta}(u_k)}{\partial a_R^2}=4A^2\Re\{s_{1k}^{*}b\}^2\sum\limits_{i=1}^{2^{p-1}}\beta_i^2e^{-\gamma_i|b|^2}
\cosh[2\beta_iu_k].
\end{split}
\end{equation} 
Thus, 
\begin{equation}
\begin{split}
\Exp{B_k^{(11)}}\hs{0.2}=\hs{0.2}4A^2\Re\{s_{1k}^{*}b\}^2\sum\limits_{i=1}^{2^{p-1}}\beta_i^2
e^{-\gamma_i|b|^2}\Exp{\frac{\cosh[2\beta_iu_k]}{F_{\theta}(u_k)}}\hs{0.3}.
\end{split}
\end{equation}
Using the PDF for $u_k$ in~\eqref{PDF_Ui}, we obtain:
\begin{equation}
\label{Exp_cosh}
\begin{split}
\Exp{\frac{\cosh[2\beta_iu_k]}{F_{\theta}(u_k)}}=\frac{2}{\sqrt{\pi MC|b|^2}}\int\limits_{-\infty}^{\infty}\hs{1}\cosh[2\beta_it]e^{-\frac{t^2}{C|b|^2}}dt.
\end{split}
\end{equation}
Moreover, for $\alpha>0$, it can be verified that
\begin{equation}
\label{standard_integral1}
\int_{-\infty}^{\infty}e^{-\alpha t^2-2\delta t}dt=\sqrt{\frac{\pi}{\alpha}}e^{\frac{\delta^2}{\alpha}}.
\end{equation}
Hence,
\begin{equation}
\label{Exp_cosh2}
\Exp{\frac{\cosh[2\beta_iu_k]}{F_{\theta}(u_k)}}=\frac{2}{\sqrt{M}}e^{\gamma_i|b|^2}
\end{equation}
and
\begin{equation}
\Exp{B_k^{(11)}}=\frac{8A^2}{\sqrt{M}}\Re\{s_{1k}^{*}b\}^2\sum\limits_{i=1}^{2^{p-1}}\beta_i^2.
\end{equation}
Using very similar steps, we can also show that
\begin{equation}
\Exp{H_k^{(11)}}=\frac{8A^2}{\sqrt{M}}\Im\{s_{1k}^{*}b\}^2\sum\limits_{i=1}^{2^{p-1}}\beta_i^2.
\end{equation}

\section{}
\label{appendixB}

In this appendix, we prove~\eqref{Term33_1}. The second derivative of $F_{\theta}(u_k)$ with respect to $b_R$ is given by
\begin{equation}
\label{dF2Rb2}
\begin{split}
&\frac{\partial^2 F_{\theta}(u_k)}{\partial b_R^2}=\sum\limits_{i=1}^{2^{p-1}}(4\gamma_i^2b_R^2-2\gamma_i)e^{-\gamma_i|b|^2}\\
&\ \ \ \ \ \ \ \ \ \ \times \cosh[2\beta_i(b_Rx_k+b_Iy_k)]\\
&-\sum\limits_{i=1}^{2^{p-1}}8\beta_i\gamma_i b_Rx_ke^{-\gamma_i|b|^2}\sinh[2\beta_i(b_Rx_k+b_Iy_k)]\\
&+\sum\limits_{i=1}^{2^{p-1}}4\beta_i^2x_k^2e^{-\gamma_i|b|^2}\cosh[2\beta_i(b_Rx_k+b_Iy_k)].
\end{split}
\end{equation}
We let
\begin{equation}
\begin{split}
T_1\triangleq&\sum\limits_{i=1}^{2^{p-1}}(4\gamma_i^2b_R^2-2\gamma_i)e^{-\gamma_i|b|^2}\cosh[2\beta_i(b_Rx_k+b_I y_k)]\\
T_2\triangleq&-\sum\limits_{i=1}^{2^{p-1}}8\beta_i\gamma_i b_Rx_ke^{-\gamma_i|b|^2}\sinh[2\beta_i(b_Rx_k+b_I y_k)]\\
T_3\triangleq&\sum\limits_{i=1}^{2^{p-1}}4\beta_i^2x_k^2e^{-\gamma_i|b|^2}\cosh[2\beta_i(b_Rx_k+b_I y_k)].
\end{split}
\end{equation}
We next find $\Exp{\frac{T_1}{F_{\theta}(u_k)}}$, $\Exp{\frac{T_2}{F_{\theta}(u_k)}}$, and $\Exp{\frac{T_3}{F_{\theta}(u_k)}}$. Using~\eqref{Exp_cosh}, we obtain
\begin{equation}
\label{T1_1}
\begin{split}
\Exp{\frac{T_1}{F_{\theta}(u_k)}}=\frac{1}{\sqrt{M}}\sum\limits_{i=1}^{2^{p-1}}(8\gamma_i^2b_R^2-4\gamma_i).
\end{split}
\end{equation}
We next consider $\Exp{\frac{T_2}{F_{\theta}(u_k)}}$. We have
\begin{equation}
\label{T2_1}
\begin{split}
\Exp{\frac{T_2}{F_{\theta}(u_k)}}&=-\sum\limits_{i=1}^{2^{p-1}}8\beta_i\gamma_i b_Re^{-\gamma_i|b|^2}\\
&\times\Exp{\frac{x_k\sinh[2\beta_i(b_Rx_k+b_I y_k)]}{F_{\theta}(b_Rx_k+b_Iy_k)}}
\end{split}
\end{equation}
Using the PDF $f_{X,Y}(x,y)$ in~\eqref{PDF_XY}, we obtain
\begin{equation}
\label{integral_appB}
\begin{split}
&\Exp{\frac{x_k\sinh[2\beta_i(b_Rx_k+b_I y_k)]}{F_{\theta}(b_Rx_k+b_Iy_k)}}=\frac{4}{\pi MC}\times\\
&\iint\limits_{-\infty}^{\hs{-3}\infty}x\sinh[2\beta_i(b_Rx+b_I y)]F_{\theta}(b_Ix-b_Ry)e^{-\frac{x^2+y^2}{C}}dxdy\\
&=\frac{1}{\pi MC}\sum\limits_{\ell=1}^{2^{p-1}}e^{-\gamma_{\ell}|b|^2}\hs{0.5}\iint\limits_{-\infty}^{\hs{-3}\infty}\hs{0.5}x\left(e^{2\beta_i(b_Rx+b_I y)}-e^{-2\beta_i(b_Rx+b_I y)}\right)\\
&\ \ \ \ \ \ \ \ \ \ \times\left(e^{2\beta_{\ell}(b_Ix-b_R y)}+e^{-2\beta_{\ell}(b_Ix-b_R y)}\right)e^{-\frac{x^2+y^2}{C}}dxdy\\
&=\frac{1}{\pi MC}\sum\limits_{\ell=1}^{2^{p-1}}e^{-\gamma_{\ell}|b|^2}\Bigg[\int\limits_{-\infty}^{\infty}xe^{-\frac{x^2}{C}+(2\beta_ib_R+2\beta_{\ell}b_I)x}dx\\
&\ \ \ \ \ \ \ \ \ \ \ \ \times\int\limits_{-\infty}^{\infty}e^{-\frac{y^2}{C}+(2\beta_ib_I-2\beta_{\ell}b_R)y}dy\\
&+\int\limits_{-\infty}^{\infty}xe^{-\frac{x^2}{C}+(2\beta_ib_R-2\beta_{\ell}b_I)x}dx\int\limits_{-\infty}^{\infty}e^{-\frac{y^2}{C}+(2\beta_ib_I+2\beta_{\ell}b_R)y}dy\\
&-\int\limits_{-\infty}^{\infty}xe^{-\frac{x^2}{C}-(2\beta_ib_R-2\beta_{\ell}b_I)x}dx\int\limits_{-\infty}^{\infty}e^{-\frac{y^2}{C}-(2\beta_ib_I+2\beta_{\ell}b_R)y}dy\\
&-\int\limits_{-\infty}^{\infty}xe^{-\frac{x^2}{C}-(2\beta_ib_R+2\beta_{\ell}b_I)x}dx\int\limits_{-\infty}^{\infty}e^{-\frac{y^2}{C}-(2\beta_ib_I-2\beta_{\ell}b_R)y}dy\Bigg].
\end{split}
\end{equation}
In the above expression, the single integrals in $x$ can be evaluated using the following result which holds for $\alpha>0$~\cite{abramowitz}:
\begin{equation}
\label{standard_integral2}
\int\limits_{-\infty}^{\infty}te^{-\alpha t^2-2\delta t}dt=-\sqrt{\frac{\pi}{\alpha^3}}\delta e^{\frac{\delta^2}{\alpha}},
\end{equation}
while the single integrals in $y$ can be evaluated using~\eqref{standard_integral1}. After evaluating all the single integrals in~\eqref{integral_appB}, we finally obtain
\begin{equation}
\label{T2_2}
\Exp{\frac{x\sinh[2\beta_i(b_Rx+b_I y)]}{F_{\theta}(b_Rx+b_Iy)}}=\frac{4C}{M}2^{p-1}e^{\gamma_i|b|^2}\beta_ib_R.
\end{equation}
Therefore
\begin{equation}
\label{T2_3}
\Exp{\frac{T_2}{F_{\theta}(u_k)}}=-\frac{16}{\sqrt{M}}\sum\limits_{i=1}^{2^{P-1}}\gamma_i^2b_R^2.
\end{equation}
We next consider $\Exp{\frac{T_3}{F_{\theta}(u_k)}}$. We have
\begin{equation}
\label{T3_1}
\begin{split}
\Exp{\frac{T_3}{F_{\theta}(u_k)}}&=\sum\limits_{i=1}^{2^{p-1}}4\beta_i^2e^{-\gamma_i|b|^2}\times\\
&\Exp{\frac{x_k^2\cosh[2\beta_i(b_Rx_k+b_I y_k)]}{F_{\theta}(b_Rx_k+b_Iy_k)}}.
\end{split}
\end{equation}
To evaluate~\eqref{T3_1}, we can follow the same approach that we used in~\eqref{integral_appB}.
However, instead of using~\eqref{standard_integral2}, we would use the following result, which holds for $\alpha>0$~\cite{abramowitz}:
\begin{equation}
\label{standard_integral3}
\int\limits_{-\infty}^{\infty}t^2e^{-\alpha t^2-2\delta t}dt=\sqrt{\frac{\pi}{\alpha^5}}\delta^2e^{\frac{\delta^2}{\alpha}}+\frac{1}{2}\sqrt{\frac{\pi}{\alpha^3}}e^{\frac{\delta^2}{\alpha}}.
\end{equation}
After some calculations (the details are skipped for brevity) we obtain 
\begin{equation}
\label{integral_appB2}
\begin{split}
&\Exp{\frac{x_k^2\cosh[2\beta_i(b_Rx_k+b_I y_k)]}{F_{\theta}(b_Rx_k+b_Iy_k)}}=\\
&e^{\gamma_i|b|^2}\bigg(\frac{C}{\sqrt{M}}+\frac{2C}{\sqrt{M}}\gamma_ib_R^2+\frac{4C}{M}\sum\limits_{\ell=1}^{2^{p-1}}\gamma_{\ell}b_I^2\bigg).
\end{split}
\end{equation}
Hence,
\begin{equation}
\label{T3_2}
\begin{split}
\Exp{\frac{T_3}{F_{\theta}(u_k)}}&=\frac{4}{\sqrt{M}}\sum\limits_{i=1}^{2^{p-1}}\gamma_i+\frac{8}{\sqrt{M}}\sum\limits_{i=1}^{2^{p-1}}\gamma_i^2b_R^2\\
&+\frac{16}{M}\sum\limits_{i=1}^{2^{p-1}}\sum\limits_{\ell=1}^{2^{p-1}}\gamma_i\gamma_{\ell}\Im\{b\}^2.
\end{split}
\end{equation}

Adding~\eqref{T1_1},~\eqref{T2_3} and~\eqref{T3_2}, we finally obtain
\begin{equation}
\Exp{B_k^{(33)}}=\frac{16}{M}\sum\limits_{i=1}^{2^{p-1}}\sum\limits_{\ell=1}^{2^{p-1}}\gamma_i\gamma_{\ell}\Im\{b\}^2.
\end{equation}
Following very similar steps, it can be shown that $\Exp{H_k^{(33)}}=\Exp{B_k^{(33)}}$.

\section{}
\label{appendixC}
In this appendix, we prove~\eqref{EqBk55}. Taking the  derivative of $F_{\theta}(u_k)$ twice with respect to $\tau$, we obtain:
\begin{equation}
\label{appC0}
\begin{split}
&\frac{\partial^2 F_{\theta}(u_k)}{\partial\tau^2}=\sum\limits_{i=1}^{2^{p-1}}\left(A^4\sigma^4\beta_i^4|b|^4-2\frac{A^4\sigma^4}{C}\beta_i^2|b|^2\right)\\
&\hs{-15}\times e^{-\gamma_i|b|^2}\cosh[2\beta_iu_k]\\
&+\sum\limits_{i=1}^{2^{p-1}}\left(4\frac{A^4\sigma^4}{C^2}\beta_i-4\frac{A^4\sigma^4}{C}\beta_i^3|b|^2\right)u_ke^{-\gamma_i|b|^2}\sinh[2\beta_iu_k]\\
&+4\sum\limits_{i=1}^{2^{p-1}}\frac{A^4\sigma^4}{C^2}\beta_i^2u_k^2e^{-\gamma_i|b|^2}\cosh[2\beta_iu_k].
\end{split}
\end{equation}
Since $B_k^{(55)}=\frac{\frac{\partial^2 F_{\theta}(u_k)}{\partial\tau^2}}{F_{\theta(u_k)}}$, it is clear from~\eqref{appC0} that we need to evaluate the terms $\Exp{\frac{\cosh[2\beta_iu_k]}{F_{\theta}(u_k)}}$, $\Exp{\frac{u_k\sinh[2\beta_iu_k]}{F_{\theta}(u_k)}}$ and $\Exp{\frac{u_k^2\cosh[2\beta_iu_k]}{F_{\theta}(u_k)}}$. The first term is found in~\eqref{Exp_cosh2}. For the second term, we have
\begin{equation}
\Exp{\frac{u_k\sinh[2\beta_iu_k]}{F_{\theta(u_k)}}}=\frac{2}{\sqrt{\pi MC|b|^2}}\hs{0.5}\int\limits_{-\infty}^{\infty}\hs{0.5}t\sinh[2\beta_it]e^{-\frac{t^2}{C|b|^2}}dt.
\end{equation}
The above integral can be evaluated using the result in~\eqref{standard_integral2}, thus obtaining
\begin{equation}
\label{appC1}
\Exp{\frac{u_k\sinh[2\beta_iu_k]}{F_{\theta(u_k)}}}=\frac{2}{\sqrt{M}}C|b|^2e^{\gamma_i|b|^2}\beta_i.
\end{equation}
For the third term, we have
\begin{equation}
\Exp{\frac{u_k^2\cosh[2\beta_iu_k]}{F_{\theta}(u_k)}}=\frac{2}{\sqrt{\pi MC|b|^2}}\hs{0.5}\int\limits_{-\infty}^{\infty}\hs{0.5}t^2\sinh[2\beta_it]e^{-\frac{t^2}{C|b|^2}}dt.
\end{equation}
We evaluate the above integral using~\eqref{standard_integral3}, obtaining
\begin{equation}
\label{appC2}
\Exp{\frac{u_k^2\cosh[2\beta_iu_k]}{F_{\theta}(u_k)}}=\frac{1}{\sqrt{M}}e^{\gamma_i|b|^2}(2C^2|b|^4\beta_i^2+C|b|^2).
\end{equation}
Using~\eqref{Exp_cosh2},~\eqref{appC1} and~\eqref{appC2}, we finally see that
\begin{equation}
\Exp{B_k^{(55)}}=\sum\limits_{i=1}^{2^{p-1}}\frac{2A^4\sigma^4}{\sqrt{M}}\left(\beta_i^4|b|^4+\frac{4}{C}\beta_i^2|b|^2\right).
\end{equation}

The above derivation can be replicated to obtain $\Exp{H_k^{55}}$ by replacing $u_k$ with $v_k$, which means that $\Exp{H_k^{(55)}}=\Exp{B_k^{(55)}}$.

\section{}
\label{appendixD}
In this appendix, we show that $\Exp{\frac{\partial^2\mathcal{L}(\bsm{q};\bsm{\theta})}{\partial a_R\partial a_I}}=0$. We have
\begin{equation}
\label{exp_RaIa}
\begin{split}
\Exp{\frac{\partial^2\mathcal{L}(\bsm{q};\bsm{\theta})}{\partial a_R\partial a_I}}&=\sum\limits_{k=1}^{N}\Exp{B_k^{(12)}-G_k^{(12)}}\\
&+\sum\limits_{k=1}^{N}\Exp{H_k^{12}-W_k^{(12)}}.
\end{split}
\end{equation}
To prove that $\Exp{\frac{\partial^2\mathcal{L}(\bsm{q};\bsm{\theta})}{\partial a_R\partial a_I}}=0$, we will show that $\Exp{B_k^{(12)}}=-\Exp{H_k^{(12)}}$ and $\Exp{G_k^{(12)}}=-\Exp{W_k^{(12)}}$.
We have
\begin{equation}
\label{appD1}
\begin{split}
\frac{\partial^2 F_{\theta}(u_k)}{\partial a_R\partial a_I}=4A^2\Re\{s_{1k}^{*}b\}\Im\{s_{1k}^{*}b\}&\sum\limits_{i=1}^{2^{p-1}}\beta_i^2e^{-\gamma_i|b|^2}\\
&\times\cosh[2\beta_iu_k]
\end{split}
\end{equation}
and
\begin{equation}
\label{appD2}
\begin{split}
\frac{\partial^2 F_{\theta}(v_k)}{\partial a_R\partial a_I}=-4A^2\Re\{s_{1k}^{*}b\}\Im\{s_{1k}^{*}b\}&\sum\limits_{i=1}^{2^{p-1}}\beta_i^2e^{-\gamma_i|b|^2}\\
&\times\cosh[2\beta_iv_k].
\end{split}
\end{equation}
Hence,
\begin{equation}
\label{appD3}
\begin{split}
\Exp{B_k^{(12)}}=&4A^2\Re\{s_{1k}^{*}b\}\Im\{s_{1k}^{*}b\}\sum\limits_{i=1}^{2^{p-1}}\beta_i^2e^{-\gamma_i|b|^2}\\
&\times\Exp{\frac{\cosh[2\beta_iu_k]}{F_{\theta}(u_k)}},
\end{split}
\end{equation}
and
\begin{equation}
\label{appD4}
\begin{split}
\Exp{H_k^{(12)}}=&-4A^2\Re\{s_{1k}^{*}b\}\Im\{s_{1k}^{*}b\}\sum\limits_{i=1}^{2^{p-1}}\beta_i^2e^{-\gamma_i|b|^2}\\
&\times\Exp{\frac{\cosh[2\beta_iv_k]}{F_{\theta}(v_k)}}.
\end{split}
\end{equation}
Since $u_k$ and $v_k$ are i.i.d., it is clear that $\Exp{B_k^{(12)}}=-\Exp{H_k^{(12)}}$. 

We next consider $\Exp{G_k^{(12)}}$ and $\Exp{W_k^{(12)}}$. To obtain $\Exp{G_k^{(12)}}$ we need $\frac{\partial F_{\theta}(u_k)}{\partial a_R}$ which is given in~\eqref{appA1} and we also need $\frac{\partial F_{\theta}(u_k)}{\partial a_I}$, which is given by
\begin{equation}
\label{appD5}
\begin{split}
\frac{\partial F_{\theta}(u_k)}{\partial a_I}=-2A\Im\{s_{1k}^{*}b\}\sum\limits_{i=1}^{2^{p-1}}\beta_ie^{-\gamma_i|b|^2}
\sinh[2\beta_i u_k].
\end{split}
\end{equation}
Hence,
\begin{equation}
\label{appD6}
\begin{split}
G_k^{(12)}&=\frac{\frac{\partial F_{\theta}(u_k)}{\partial a_R}\frac{\partial F_{\theta}(u_k)}{\partial a_I}}{F_{\theta}(u_k)^2}\\
&=4A^2\Re\{s_{1k}^{*}b\}\Im\{s_{1k}^{*}b\}\Bigg(\frac{\sum\limits_{i=1}^{2^{p-1}}\beta_i\sinh[2\beta_iu_k]}{F_{\theta}(u_k)}\Bigg)^2.
\end{split}
\end{equation}

For $W_k^{(12)}$, we need $\frac{\partial F_{\theta}(v_k)}{\partial a_R}$ and $\frac{\partial F_{\theta}(v_k)}{\partial a_I}$, which are respectively given by
\begin{equation}
\frac{\partial F_{\theta}(v_k)}{\partial a_R}=-2A\Im\{s_{1k}^{*}b\}\sum\limits_{i=1}^{2^{p-1}}\beta_ie^{-\gamma_i|b|^2}
\sinh[2\beta_i v_k],
\end{equation}
and
\begin{equation}
\frac{\partial F_{\theta}(v_k)}{\partial a_I}=2A\Re\{s_{1k}^{*}b\}\sum\limits_{i=1}^{2^{p-1}}\beta_ie^{-\gamma_i|b|^2}
\sinh[2\beta_i v_k].
\end{equation}
Thus,
\begin{equation}
\label{appD7}
\begin{split}
W_k^{(12)}&=\frac{\frac{\partial F_{\theta}(v_k)}{\partial a_R}\frac{\partial F_{\theta}(v_k)}{\partial a_I}}{F_{\theta}(v_k)^2}\\
&=-4A^2\Re\{s_{1k}^{*}b\}\Im\{s_{1k}^{*}b\}\Bigg(\frac{\sum\limits_{i=1}^{2^{p-1}}\beta_i\sinh[2\beta_iv_k]}{F_{\theta}(v_k)}\Bigg)^2.
\end{split}
\end{equation}
Since $u_k$ and $v_k$ are i.i.d., it is clear from~\eqref{appD6} and~\eqref{appD7} that $\Exp{G_k^{(12)}}=-\Exp{W_k^{(12)}}$, which completes the proof.

\bibliographystyle{IEEEtran}
\bibliography{IEEEabrv,CRB_MQAM_bib}

\end{document}